# A MANGANISE IONS GROUND STATE IN $Mn_xSi_{1-x}$ : NEGATIVE - U PROPERTIES CENTRE ?

Yakubenya S.M.

RRC "Kurchatov Institute", 123182, Moscow, Russia

e-mail: benyajamp@yahoo.com

**Abstact.** The properties of magnetic deluted and strong correlated systems of $Mn_xSi_{1-x}$ systems are discussed. In frame work of the model of the double defects including manganese ion and silicon vacancy are considered properties of these systems. The role of the Jahn -Teller distortions of different symmetry types in MnSi system magnetic-properties formation is discussed. It has been established that defect $\{Mn_s^{+1} - V_{si}^{o}\}$ is the center with negative-U properties and Jahn-Teller's full symmetric vibration mode initiates change of a crystal-field value from intermediate (a configuration - $\{Mn_s^{+1} - V_{si}^{o}\}$) to strong (a configuration -$\{Mn_s^{-1} - V_{si}^{+2}\}$).

## Introduction

Intermetallic compounds based on the elements with incompletely filled 3d- or 4f-shell attracts attention of researchers on an extent more than 40 years [1÷3]. Such interest is caused firstly by all their unusual magnetic and thermodynamic properties [4,5], which directly connected with electronic structure of a magnetic ions forming corresponding sublattice of intermetallic compound. In a case of strongly diluted systems, when magnetic ions are accompanying impurity, interest to their properties is connected to their key role in formation of practically important parameters, in particular free-carrier lifetime in the system [6,7].

The problem of a manganese ions ground state in $Mn_xSi_{1-x}$ is discussed in the present work. Two limiting cases are considered: 1) a concentration manganese ions is infinitesimal as a free carriers concentration (x →0). Interaction between magnetic ions and free carrier is absent; 2) concentration of Mn is close to one of Si (x ~ 0.5). The interaction of magnetic moments manganese ions with quasiparticals is very important component in magnetic-property-compound formation. In the first case Mn ions can be treat as a doping impurity of Si crystal.

Both systems are crystallized in the cubic lattice type. Symmetry group $T_d$ is realized for the magnetic deluted system [8] and B20 take place in the case of strong correlated system correspondingly [9]. Take into account that interaction between atomic orbitals and crystal field is dominated with respect to other interactions in crystal we shall look for the electronic configuration change of the manganese ions ground state as the magnetic component-concentration increase in compound.

Magnetic moment value of the manganese ions in the systems in a wide temperature interval will be main subject of our investigation. The model of the double defect, formulated early by us [10,11] will be used for description of magnetic transformation taking place as for magnetic diluted system as for strong correlated system $Mn_{0.5}Si_{0.5}$. Shot description of the double defect model will be done below.

**Model.**

Two different defect types in semiconductors irrelevant with each to other are discussed as a usual [8]. It is named by interstitial defects -$\{X_i\}$ (impurity ions located in a crystal lattice pore) and substitution defects $\{X_s\}$ (impurity ions located in point of lattice).

In the same times, in the frame of the double defect model, the interstitial and substitution defects are considered as a two limiting cases of double defect. A silicon vacancy $\{V_{si}\}$ acts as a partner for the Mn impurity ions $\{Mn_s\}$ in the suggested double defect.

At that when,
--- distance between partners is infinity, then interstitial defect type is realized;
--- distance between partners is an infinitesimal value, then we have substitution type of defect.

The account of silicon vacancy at interstitial defect case is necessary to consider all type of double defects from the common base. This will result to unique parameter – Fermi level position ($E_F$), which will define a degree of one-electronic orbital fillings of such kind of double defect. ($E_F$) is counted from vacuum level position in the case.

The intermediate case, when distance between partners is of the order of the lattice period, the only one valence bond saturated exists between partners. Such type of defects has been named by the pairing defects. A plenty of pairing defect different types may exist, depending on overlapping integral of wave functions partners and number of nonequivalent crystallographic positions in a lattice. Only single type of pairing defects is realized in silicon. Moreover, we can see electronic density redistribution between the central ion and ions of the nearest crystal environment as a result of temperature variation or other factors, in particular, a electronic configuration change of everyone components as result of Jahn-Teller distortions in the frame of the model .It is necessary to note, that electronic density redistribution can be realized and without a free carrier generation .The magnetic-moment average value located at manganese ion can be renormalized very strongly as free carriers appear in the system

The considering of double defect model we intentionally limit to the consideration of single elementary cell, understanding thus, that significant electronic density delocalization take place due to hybridization effects and indemnification of a charge occurs much further, than the first coordination sphere. Here and further for a designation of such defects we shall use the notation for substitution $\{Mn_s^m - V_{si}^n\}$ and $\{Mn_i^m - V_{si}^n\}$, for pairing defects and interstitial defects $\{Mn_i^m\} - \{V_{si}^n\}$. m and n indexes designate a charge of everyone components with respect to its nucleus. Take into account an electroneutrality condition of a crystal as a whole, we can write equation for the neutral charge state of double defect:

$$m + n \equiv 0 \qquad (1)$$

, where **m** and **n** can take different value, not only zero as it practises at the standard approach of the description of substitution defects behaviour in semiconductors[8]. In other words, numbers of electrons, located on valence bonds between central ion and the nearest neighbors, may be differ from one to one for the impurity ion cases and ions of the basic lattice. Four electrons occupy silicon vacancy orbitals for the $V_{si}^o$ case.

Now all are ready for proceed to discussion of a manganese ion ground state in intermetallic compound $Mn_xSi_{1-x}$ and in strongly diluted system silicon – manganese. We shall begin a discussion from the case of strongly diluted system: silicon - manganese.

**$Mn_xSi_{1-x}$ strongly diluted system ($x \to 0$)**

The crystal lattice in such compound will mainly consist of silicon atoms and is described by point group of symmetry $T_d$ [8]. Silicon demonstrates semiconductor properties and gap is in a one-partial excitation spectrum. The forbidden band value $E_g$ is equal to 1.17 eV at temperatures close to absolute zero [12].

Mono- and polycrystalline silicon used as a material for solar batteries, is presented up to 15 metal impurities at concentration above than $10^{12}$ ions/sm$^3$ and higher [13,14]. Their number includes also manganese. Transition metal impurities located in interstitial position in crystal lattice of silicon, as a role [8,13,14]. Sample preparation, in which main part impurity manganese ions located in unit of crystal lattice, is not ordinary task and we know few works only [15,16,17], in which authors have solved the given problem. Concentration of the point defects due to manganese impurity ions located in unit of crystal lattice, does not exceed $10^{15}$ ions/cm$^3$ in the crystal and any magnetic interactions as a manganese - manganese as a free carrier - magnetic ion can be neglected.

But, crystal field effects and hybridization between impurity and band states are played a dominant role in the formation of the magnetic-ions electronic structure. Manganese atoms have half-filled 3d–shell and two electrons occupy an outside 4s - shell. An electronic configuration of the $Mn^0$ looks like $3d^54s^2$. Impurity ion 3d-orbitals are split by tetrahedral-symmetry crystal field on the $t_2$- and $e_2$- orbitals (three- and double- degenerated states). The electronic configuration looks like $(e_{2\uparrow})^2(t_{2\uparrow})^3$ .( $\uparrow$ - spin projection up ; $\downarrow$ - spin projection down). The order of filling such orbitals is defined by a lot of parameters, including of a crystal field value (strong, intermediate, weak). The Fermi level position in the gap is defined by charging state of defects and their mutual concentration.

In early works directed to research the behaviour of manganese ions in silicon [18,19], key parameters of electronic structure of the defects connected to them have been determined, but interpretation of the received results was carried out within the framework of standard representations about substitution defects in semiconductors. Repeatedly, a question on electronic structure of substitution defects connected with manganese ions have returned only 30 years later, after from carrying out of the first experiments [15,16]. Let's note, that researches of EPR (Electron Paramagnetic Resonance) spectra of $Mn_xSi_{1-x}$ ($x\rightarrow 0$) samples and simultaneously with it of spectra DLTS (Deep Level Transition Spectroscopy) [15,16] have allowed to remove question on the magnetic moment value located on a magnetic ion (Fig. 1,2).

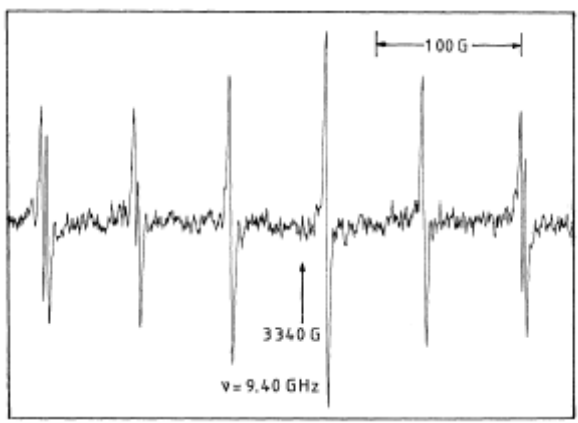
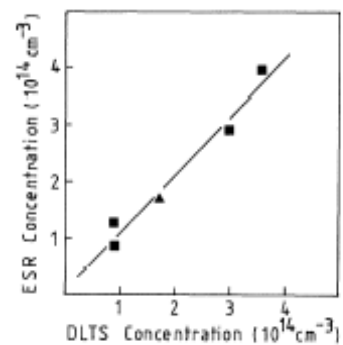

| Fig.1 | Fig.2 |
|---|---|
| EPR first-derivative absorption signal from $Mn_s^+$ in arbitrary orientation [15]. | Correlation plot of 0.38 eV DLTS peak intensity and the integrated ESR intensity of the $Mn_s^+$ signal [15]. |

Fermi level ($E_F$) is located near the valence band top in the investigated crystals at low temperatures. It is caused by presence of shallow acceptors at concentration exceeding the general concentration of manganese ions [8]. As the result, the dominate part of manganese ions, should be in charge state $\{Mn_s^{+1} - V_{si}^{o}\}$ and the magnetic moment of manganese ions has pure spin origin. It is defended by two electrons, which occupied $e_2$-orbitals (configuration $(e_{2\uparrow})^2$ ). Moreover, it is expected, what other electrons of manganese and silicon participate in valence bond formation, as it is take place in perfect silicon crystal. There is very critical contradiction in last postulation.

As very well know, that $V_{si}^{+2}$ is the greatest possible charging state silicon vacancy[20,21]. In other words, two electrons occupy of 3s- orbitals of silicon at any Fermi level position in the gap. Energy ionization 3s-shell of silicon is ~ 35 eV [22]. In same times, defect $\{Mn_s^{+1} - V_{si}^{o}\}$, having four saturated valence bonds between surrounding and central ions, It can be realized if $Mn^{+4}$ charge state take place. It is equivalent to postulate, that electrons with bonding energy ~ 100 eV, participate in valence bond formation. But it is not realistic suggestion.

In the same time EPR spectra of $\{Mn_s^{+1} - V_{si}^{o}\}$ defects is described by means of cubic spin-Hamiltonian very well [15,23] . At the standard approach, when one considers only electrons located on a magnetic ion, it is not look possible to find the reasonable decision of the arising contradiction.

The similar situation took place in system GaAs:Mn and was a subject of discussions more than 30 years [24÷26]. The decision of the problem within the framework of double defect model has been found some years ago [27].

Let's consider, that in valence bond formation between magnetic ion and surrounding only electrons with bonding energy in free atom less then 35 eV on absolute value are involved. Energy scale is counted from a vacuum level position. In other words, two from four valence bond is saturated in $\{Mn_s^{+1} - V_{si}^{o}\}$ defects. Two valence bond is broken. The defect in such configuration is unstable with respect to Jahn-Teller distortion. The most probable direction of distortions is directions of {110} type (tetragonal distortions of local symmetry – E-mode of vibration [28,29]). Depending on a relation between phonon local energy ($\hbar\omega_{loc}$) and depth of a potential pit ($E_{J-T}$) Jahn-Teller effect can be static or dynamic (Fig.3b,3a).

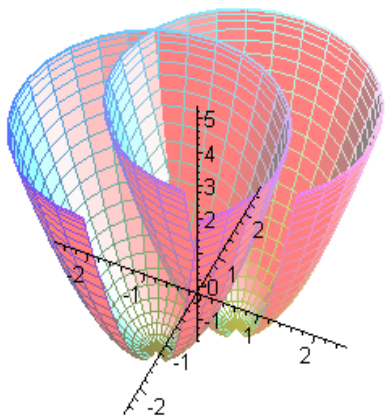 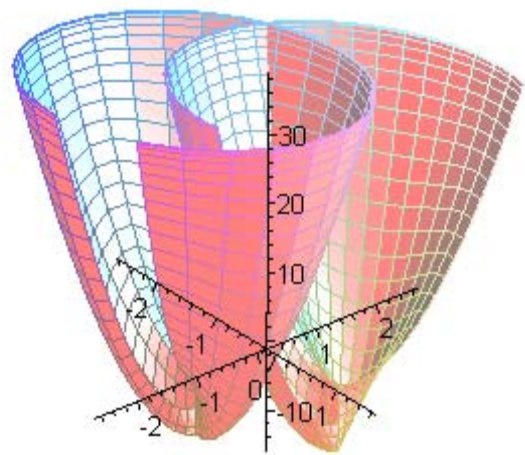

Fig.3a                                                                              Fig.3b
Potential energy surface for the                                  Potential energy for the dynamic Jahn-Teller effect case.  static Jahn-Teller effect case.
(in generalized coordinates space)  (in generalized coordinates space)

. $E_{J-T}$ and $\Delta$ - barrier value between potential energy minimums are defined as:

$$E_{J-T} = (B_1)^2/(2k_e - 4B_2) \quad (2a)$$

$$\Delta = 4B_2 * E_{J-T}/(k_e - 2B_2) \quad (2b)$$

where $B_1$, $B_2$, $k_e$ - constant (for more detail see [28]).

A dynamic type of Jahn-Teller effect takes place for the manganese ions in such system (see EPR spectra). But, we shall have full magnetic moment $J$, located on manganese ion incoming in $\{Mn_s^{+1} - V_{si}^o\}$ defect, equal two at presence tetragonal Jahn-Teller distortions only. An electronic configuration of the magnetic electrons is ( $(e_{2\uparrow})^2(t_{2\uparrow})^3(t_{2\downarrow})^1$ ). Here, $t_2$ –orbitals is a gibrids of 3p-orbitals of silicon and 3d-orbitals of manganese.

Provided that simultaneously with tetragonal distortion it takes place also full symmetric Jahn-Teller distortions of local symmetry ( $A_1$ – mode of vibrations). The full magnetic moment (J) located on an ion of manganese is equal $(\sqrt{2})\mu_B$ ($\mu_B$ - Born magneton). Electronic configuration is $(e_{2\uparrow})^2(t_{2\uparrow})^3(t_{2\downarrow})^3$. J is connected with manganese spin ( S=1 ) by the following ratio:

$$J = (\sqrt{S*(S+1)})\mu_B \quad (3)$$

As we see, very good consent between calculated and experimentally values of magnetic moment $J$ takes place. In other words, the defect $\{Mn_s^m - V_{si}^n\}$ can exist in two isoelectronic configurations:

$$m = +1 \quad n = 0 \quad (4a)$$

$$m = -1 \quad n = +2 \quad (4b)$$

but, transition between configurations take place without change of density of free carriers of a charge.

The defects with an electronic configuration (4b) are responsible for observable EPR spectra. In the same time, DLTS spectra are connected with (4a) configuration of the double defect.

In this case it is easy to coordinate value of activation energy $\varepsilon_i^{exp}$ obtained in experiment (DLTS), and $\varepsilon_i^{cal}$ (calculated). Observable value will be well coordinated with the value, received by means of "spin marking method" [30]. Experimental value $\varepsilon_i^{exp}$ is equal to 0.38 eV [15] and $\varepsilon_i^{cal}$:

$$\varepsilon_i^{cal} = 3\varepsilon_o \quad (5)$$

where $\varepsilon_o$ - is energy of one electron transition in silicon. It is equal to ~ 0.13 eV [20,21].

The process of electron capture on the deep level is defined by:

$$\{Mn_s^{-1} - V_{si}^{+2}\} + 2\varepsilon_o \Rightarrow \{Mn_s^{+1} - V_{si}^o\} + \varepsilon_o \Rightarrow \{Mn_s^o - V_{si}^o\} + h^+ \quad (6)$$

where $h^+$ - hole, the free carrier of a charge.

Within the framework of the standard approach:

$$\varepsilon_i^{cal} = \varepsilon_o \quad (7)$$

$$Mn_s^{+1} + \varepsilon_o \Rightarrow Mn_s^o + h^+ \quad (8)$$

Look discussion in [10,11] for more detail.

In the conclusion we shall note, that full symmetric vibration mode initiates change of a crystal field value from intermediate (a configuration - $\{Mn_s^{+1} - V_{si}^o\}$) to strong (a configuration -$\{Mn_s^{-1} - V_{si}^{+2}\}$. Also $\{Mn_s^{+1} - V_{si}^o\}$ defect demonstrate properties, characteristic for the centers with negative energy of coulomb correlation (centers with negative-U properties). In this connection, most likely, revision of experimental data interpretation on behaviour Fe and Ni ions in silicon is necessary.

**$Mn_xSi_{1-x}$ strongly correlated system (x ~ 0.5)**

Crystalline system MnSi, in which both components are approximately having equal concentrations, traditionally attributes to strong correlated systems [1]. As opposed to strongly deluted systems, considered above, manganese ions build up it's own sublattice and any shallow acceptors is absent in the system. Take into account, that electrical resistivity of the manganese silicite samples drop monotonically down to 20 mK [31], we can postulate that conduction band states (local level in the gap in magnetic-diluted-system case) is not empty in the system. It is formed by mean of hybridization 3p-orbitals of silicon and 4s- 3d- orbitals manganese [32].

At the same time, small distinction in types of a crystal lattice between magnetic deluted system (group of symmetry $T_d$ - four equivalent atoms of silicon in the first coordination sphere – 2.35 Å [8]) and crystals of manganese silicite (type of lattice B20 - three equivalent atoms of silicon on distance 2.396 Å and one -2.313 Å [9] at [111] direction) takes place.

Displacement of atoms of the nearest environment in the opposite direction in lattice S20 in comparison with atoms position in tetrahedral lattice is powerful argument for the existence of the cooperative Jahn – Teller effect in such system. But, distances between the nearest ions of manganese in $Mn_{0.5}Si_{0.5}$, which do not form valence bonds with each other, practically coincides with such parameter for metallic γ- manganese [33].

Systematic measurements of the magnetic moment, located on the manganese ions, show presence of several temperature intervals in which magnetic ions demonstrate completely different properties [33,34]. It is paramagnetic at temperature above 30 K. First-order transition takes place at 29.5K[5,35]. Two different types of magnetic ordering are realized at temperature below 29.5 K. Helical magnetic structure with a long-range order was observed at $H_{ext} \to 0$ ( $H_{ext}$ - external magnetic field ) and the structure becomes progressively more conical at increase of $H_{ext}$ value. Moreover, there is critical value of $H_{ext}^o$ (T) when ferromagnetic order type is realized in the system. (Fig.4).

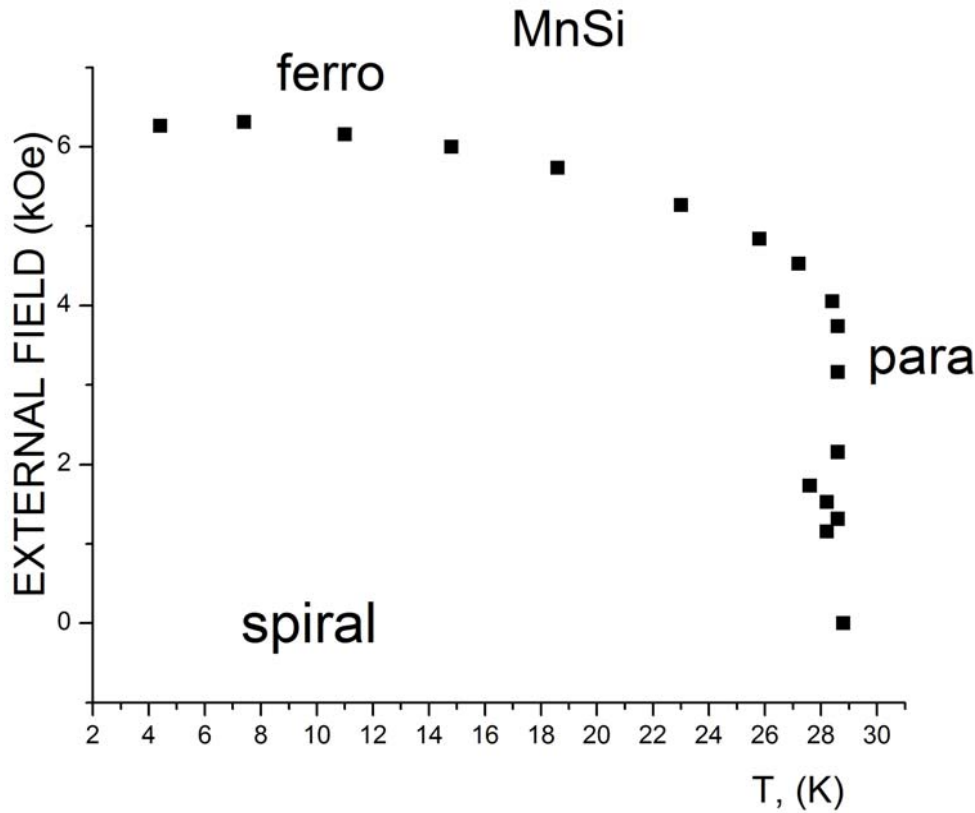

Fig.4
Magnetic phase diagram of MnSi. Inset show the
Projection of Mn atomic positions on a <100>
plane. This atomic arrangement lacks a center
of symmetry. [34]

   Within the framework of double defect model the given transformations can be treated in the following way. The value of the magnetic moment, located on the manganese ions, in a paramagnetic phase ($\mu_{para} \sim 1.4\mu_B$ [34,36,37], *g-factor is not equal 2.0 but 2.81*) practically coincides with those one for a case of magnetic deluted system. That is why we can postulate, that interaction between free carriers of charge and the magnetic moment, located on ions of manganese, can be neglected. Manganese ions precess around the [111] axes of the crystal. Cooperative Jahn-Teller effect takes place (see text above).
   First-order transition take place at 29.5 K and transition $\{Mn_s^{+1} - V_{si}^o\} \Rightarrow \{Mn_s^{+1-d} - V_{si}^{o+d}\}$ is realized. Here parameter d >0 and it depicts a redistribution electron density between partners in the double defect (see discussion for the magnetic deluted system). Let's note, that the amplitude of full symmetric vibrations in this case may be less then in magnetic deluted system. The system is closed in one of Jahn-Teller's minimum (Fig.3b). Moreover, we have got fist-order transition.
   It is necessary to note, that very intricate case of Jahn-Teller distortions is realized here – trigonal + tetragonal + full symmetrical types of distortions take place simultaneously and we have got a problem with large quantity of independent parameters. As result, numbers of Jahn-

Teller's minima can be equal to 24 [28,29]. In the case, the spiral structure is connected with a jump between minima of potential energy. As result of applied external magnetic field, the transformation of a local minimum to absolute one is a case of the transition from the spiral structure to ferromagnetic order type in $Mn_{0.5}Si_{0.5}$. Parameter $\mu_{fer}H^{o}_{ext}(T \to 0)$ can be defined, approximately, as a potential barrier $\Delta$ [29] between different Jahn-Teller's minima:

$$\Delta = \mu_{fer}H^{o}_{ext}(T \to 0) \quad (9)$$

$\mu_{fer} \cong 0.4\ \mu_B$ and $H^{o}_{ext}(T \to 0) \cong 6.5$ kOe at temperatures below $T_N$ [34]. The discussion about $\mu_{fer}$ value in frame of fluctuated Fermi-liquid model may be find in work [38].

Separately it is necessary to focus on the dependence of the phase-transition temperature from hydrostatic pressure value.

As we already discussed above, the first order phase transition, occurring at temperature 29.5 K at normal pressure, is connected with redistribution of electronic density between manganese and surrounding silicon atom. But 3p-orbitals of silicon is more delocalizated in comparison with 3d-orbitals of manganese [22]. As result:

$$(\partial E_{3p}/\partial P)_{Si} > (\partial E_{3d}/\partial P)_{Mn} \quad (10)$$

where P - hydrostatic pressure value.

But if ratio (10) is corrected, the phase transition temperature has to drop at the hydrostatic-pressure increase. Moreover, there is critical value $P_{critical}$ when transition is impossible. $\{Mn_s^{-1} - V_{si}^{+2}\}$ defect configuration is unfavorable and phase transition is not observed. It was confirmed experimentally [5,31].

**Conclusion**

Thus, the model of double defect, stated above, can be successfully applied for the description of a magnetic properties of $Mn_xSi_{1-x}$ system with different concentration of manganese ions. It is assume, that local level, located in the forbidden band of silicon and connected with recharging of impurity manganese ions, transforms in the band with increasing of manganese concentration up to x ~ 0.5 in the $Mn_xSi_{1-x}$ system. We found correct descriptions for the level position in the gap of silicon at low manganese concentration in frame of double defect model. The transition from intermediate crystal field to strong one as the result of full symmetry of Jahn-Teller vibration mode has been detected.

It has been established, that $\{Mn_s^{+1} - V_{si}^{o}\}$ defect is the center with negative-U properties.

The favorable difference of our model from the approach of weakly-fluctuated Fermi-liquid one, which widely used for the description of physical properties of manganese silicite in last years [38,39], is the opportunity of definition of original causes of those or other effects.

It is obvious, that is necessary perform a number of experimental and theoretical studies with the aim of the specification of our model parameters. First of all, it concerns measurement of inelastic neutrons scattering spectra on the MnSi single-crystalline sample and on amorphous a-MnSi, in which helical magnetic order is absent. It is fruitful to compare EPR spectra of $Mn_xSi_{1-x}$ ($x \to 0$) and a-MnSi in a wide temperature interval.

Useful addition to these experiments would be a comparison of the μ+-mesons scattering experiment results and Knight's shift in spectra of nuclear magnetic resonance MnSi at temperatures ~ $T_N$.

**Acknowledges**

The author is indebted to P.A.Alekseev and V.N.Luzukov for helpful discussion and critical reading of the manuscript.

**References**

[1]  U.A. Izumov et.al.: Uspekhi Fizicheskikh Nauk (in Russian) ,v.178,(2008),  p25

[ 2]  P.A.Alekseev  et.al: J.Phys. :Condens.Matter v.16,(2004), p.2631

[3]    R.Kadano et al. :     Phys.Rev.v.42B,(1990), p.6515

[4]    S.V.Grigoriev et.al : Phys.Rev.v.73B,(2006), p.224440

[5]    A.E.Petrova et.al. : Phys.Rev.v.74B,(2006), p.092401

[6]    D.Macdonald et.al. :   Semicond.Sci.Technol.v.22, (2007 ),  p.163

[7]    Roth et. al .: J.Appl. Phys.v.102 (2007) , 103716

[8] W.A.Harrison : Electronic structure and the properties of solids.(San Francisco, Freeman,1980)

[9]   T.M.Hayes  et. al. :   Phys.Rev.v.23B,(1981), p.4691

[10] S.M. Yakubenya: Fizika Tverdo Tela ( in Russian) v.33,(1991), p.1462

[11] S.M. Yakubenya: :Fizika Tverdo Tela (in Russian)v.33,(1991), p.1470

[12] I.S. Grigor'ev et al: Handbook  Physical values ( in Russian),  Enorgoatomizdat (1991)

[13] T.Buonassisi et. al .: Acta Materialia v.55,(2007), p.6119

[14] A.A.Istratov et.al.: Applied Physics A, v.A69, (1999), p.13

[15]  R.Czaputa et .al.: Phys.Rev.Lett.,v.55 (1985),  p.758

[16]  M.Haider et.al. : J.Appl. Phys.v.62 (1987), p.3785

[17]  K.Graff :Metal Impurities in Silicon- Devise Fabrication, v.24, 2nd ed.( Springer, Berlin,1999)

[18]  R.O.Carlson: Phys.Rev.,v.104,(1956), p.937

[19]  H.Ludwig et. al., Electronic Paramagnetic resonance in Solid State Physics,v.13 (Academic

,NY,1962)

[20] G.A.Baraff et. al: Phys. Rev., v.21B,(1980), p.5662

[21] Newton et.al.: Physica B+C , v.116B, (1983), p.219

[22] A.A.Radtsig et.al. Handbook Parameters of atoms and atomic ions , Energoatomizdat,(in Russian), 2nd ed.(1986)

[23] A.Abragam et.al., Electron Paramagnetic Resonance of Transition Ions,v1,2 (Clarendon Press, Oxford , 1970)

[24] D.G. Andrianov et. al. Fizika i Tekhnika Poluprovodnikov (in Russian),v.17 ,(1983), p.810

[25] A.E.Vasil'ev et.al., Fizika i Tekhnika Poluprovodnikov (in Russian),v.17 ,(1983), p.1823

[26] J.Schneider et.al., : Phys.Rev.Lett.,v.59 ,(1987), p.240

[27] S.M.Yakubenya : J.Moscow Phys.Soc., v.7, (1997), p.273

[28] I.B.Bersuker The Jahn_Teller Effect (NY, Plenum, 1983)

[29] I.B. Bersuker et.al., Uspekhi Fizicheskikh Nauk (in Russian) ,v.116,(1975), p605

[30] S.M. Yakubenya : Materials Science Forum, v.83 - 87 , (1991), p. 363

[31] C.Pfleiderer et. al. : : Phys.Rev.,v.55B,(1997), p.8330

[32] S.-J.Oh et.al., : Phys.Rev.,v.35B,(1987), p.2267

[33] Y.Ishikawa et.al. :   Phys.Rev.v.16B,(1977) ,p.4956

[34] G.Shirane et.al. :  Phys.Rev.,v.28B,(1983) , p.6251

[35] S.M.Stishov et. al.  Phys.Rev.,v.76B,(2007) ,p.052405

[36] K.R.A.Ziebeck et.al.: J.Phys.F: Metal Phys. v.10, (1980), p.2015

[37] K.R.A.Ziebeck et.al.: J.Phys.F: Metal Phys. v.11, (1981), p.L127

[38] S.V.Maleyev . Phys. Rev. ,v.73B,(2006), p. 174402

[39] D. Belitz et.al. Phys. Rev. ,v.74B,(2006), p. 024409